\documentclass[pre,aps,twocolumn]{revtex4}
\usepackage{epsfig}

\begin{document}  

\title{
Hydrodynamic correlations in the translocation of biopolymer through a nanopore:
theory and multiscale simulations 
}
\author{Maria Fyta$^1$, Simone Melchionna$^{2,3}$, 
Sauro Succi$^{4,5}$ and Efthimios Kaxiras$^{1,2}$}
\affiliation{$^1$Department of Physics and $^2$School of Engineering 
and Applied Sciences, Harvard University, Cambridge, MA, USA\\
$^3$INFM-SOFT, Department of Physics, Universit\`a di Roma 
\it{La Sapienza}, P.le A. Moro 2, 00185 Rome, Italy \\
$^4$Istituto Applicazioni Calcolo, CNR, 
Viale del Policlinico 137, 00161, Roma, Italy \\
$^5$Initiative in Innovative Computing, Harvard University,
Cambridge, MA, US
}

\date{\today}

\begin{abstract}

We investigate the process of biopolymer translocation through a 
narrow pore using a multiscale approach which
explicitly accounts for the hydrodynamic interactions of the 
molecule with the surrounding solvent. 
The simulations confirm that the coupling of the
correlated molecular motion to hydrodynamics results in 
significant acceleration of the translocation process.
Based on these results, we construct a phenomenological model
which incorporates the statistical and dynamical 
features of the translocation process and  
predicts a power law dependence of the translocation time on 
the polymer length with an exponent $\alpha \approx 1.2$.  
The actual value of the exponent from the simulations is $\alpha = 1.28 \pm 0.01$,  
which is in excellent agreement with experimental measurements of 
DNA translocation through a nanopore, and is not sensitive to the choice
of parameters in the simulation.
The mechanism behind the emergence of such a robust exponent
is related to the interplay between the longitudinal and transversal
dynamics of both translocated and untranslocated segments.
The connection to the macroscopic picture involves separating the contributions
from the blob shrinking and shifting processes, which are both essential to the
translocation dynamics.
\end{abstract}


\maketitle

\section{Introduction}

Translocation of biopolymers, such as DNA and RNA, plays a vital role
in many important biological processes, such as
viral infection by phages, inter-bacterial DNA transduction
or gene therapy \cite{TRANSL}. This has motivated a number 
of {\it in vitro} experimental studies, aimed at exploring the 
translocation process through protein channels across cellular
membranes \cite{EXPRM1,EXPRM2}, or through 
micro-fabricated channels \cite{DEK}.
In particular, recent experimental work has 
focussed on the possibility of fast DNA-sequencing by ``reading-off'' 
the DNA bases while tracking its motion through nanopores under 
the effect of a localized electric field \cite{nature03,NANO}.  

The translocation of biopolymers is a complex phenomenon 
involving competition between many-body atom-atom interactions, 
fluid-atom hydrodynamic coupling, as well as the interaction of the 
polymer with wall molecules in the nanopore. Although some universal 
features of the translocation process can be analyzed by means of suitably 
simplified statistical models 
\cite{Nelson,statisTrans1,statisTrans2,statisTrans3},
and non-hydrodynamic coarse-grained or microscopic models 
\cite{DynamPRL,Kotsev-2007,numer1,numer2}, a 
quantitative description of this complex phenomenon calls for realistic, 
state-of-the-art computational modeling. 
Work along these lines has been recently reported by several groups, 
beginning with the first multiscale 
simulations by the present authors \cite{ourLBM,ourIJMPC07}, 
followed by Langevin dynamics simulations \cite{Muthukumar-2007} and
more recently by coupled molecular-fluid dynamics \cite{dePablo-2008,ourNL08}. 
Specifically, Forrey and Muthukumar performed Langevin dynamics simulations, and
examined single-file as well as multi-file translocation \cite{Muthukumar-2007}.
Izmitli {\it et al.} used a coupled lattice Boltzmann - Molecular Dynamics scheme \cite{dePablo-2008}, as outlined in 
\cite{ourLBM,ourIJMPC07}, and reproduced our early results, while exploring
a smaller range of chain lengths and ensemble size.
These recent works have provided a wealth of new computational results 
and detailed insight into the problem of translocation through nanopores.
In addition to these studies, Slater and co-workers have investigated 
the translocation process numerically by also including hydrodynamic interactions
\cite{slater}. However, these authors treat the cases were no external field is
applied, thus comparison with our work cannot be made at this point.

In this work, we report a synthesis of the simulational result into a coherent
mean-field analytical model which captures the basic physical mechanisms behind 
the translocation process.  The model is based on extracting the scaling
behavior of translocated and untranslocated segments, including the 
anisotropy between longitudinal and transverse components.  The
analytical model  
predicts a power-law scaling behavior of translocation time with polymer
length with exponent $\alpha \approx 1.2$, which is very close to
the one found in the current ($\alpha = 1.28 \pm 0.01$) and
other ($\alpha = 1.28 \pm 0.03$ \cite{dePablo-2008}) simulations and in 
experiments of DNA translocation ($\alpha = 1.27 \pm 0.03$ \cite{NANO}).

The paper is organized as follows: in Section II we provide an overview of our 
simulational approach.
In Section III we discuss the choice of simulation parameters that make our simulations
relevant to DNA translocation through nanopores and the implications for the 
implied time-scales and length-scales of the system.
In Section IV we discuss the results of the translocation simulations, paying particular
attention to the anisotropy of longitudinal and transverse components of the translocating
polymer in both the untranslocated and translocated segments.
Section V presents the analytical mean-field model and its comparison to the 
simulations.  We conclude in Section VI with some comments on what may be 
the limitations of the mean-field picture and a summary of our results.

\section{Multiscale model}

Our multiscale method is based on the coupling between
constrained Molecular Dynamics (MD) for the polymer evolution and a 
lattice Boltzmann (LB) treatment of the explicit solvent dynamics \cite{LBE1,LBE2}.
In contrast to Brownian dynamics, the LB approach 
handles the fluid-mediated solute-solute 
interactions through an explicit representation of local
collisions between the solvent and solute molecules.
We will focus on the {\it fast} translocation regime, in
which the translocation time $\tau$ is much smaller than the
typical relaxation (Zimm) time of the 
polymer towards its native (minimum energy, maximum entropy) 
configuration. 
This regime cannot be captured by a simple, one-dimensional 
Brownian model \cite{KARDAR} or a Fokker-Planck representation.

\begin{figure}
\begin{center}
\includegraphics[width=0.4\textwidth]{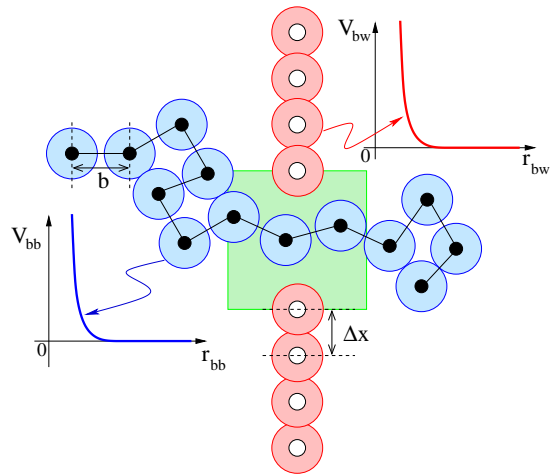}
\caption{(Color online) Illustration of the interactions in the DNA translocation model: 
the beads representing the DNA are shown as small black dots connected by straight line 
segments, the wall is represented 
by a plane of on-lattice points (small white dots separated 
by the lattice spacing $\Delta x$) which repel the DNA beads within a range of interaction. 
The bead-bead interactions are indicated by 
the blue spheres surrounding the small black ones, defined in Eq.(\ref{eq:LJ}), except for the distance 
between consecutive beads which is fixed at $b$.
Interactions between the beads and the wall are
indicated by the red spheres surrounding the small white ones,  defined in Eq. (\ref{eq:LJ}).
Interaction between
the beads and the constant external field are confined over a shaded (in green) 
region around the pore (see text for details).}
\label{LJinteractions}
\end{center}
\end{figure}

Translocation is induced by the constant electric field $\vec{E}$ 
acting along the $x$ direction localized in a region near the pore.
The dynamics of the beads which constitute the 
molecule are governed by the equation
\begin{equation}
m_b \frac{d\vec{v}_i}{dt} = 
\vec{F}_{tot,i} = \vec{F}_{c,i} + \vec{F}_{drag,i} 
+ \vec{F}_{r,i}+\vec{F}_{\kappa,i} +\vec{F}_{drive,i}
\label{eq:force}
\end{equation}
with $\vec{F}_{tot,i}$ the 
total force on bead $i$.  
$\vec{F}_{c,i}$ is a conservative force describing the sum of bead-bead 
and bead-wall interactions; 
$\vec{F}_{drag,i}$  
is the dissipative drag force due to polymer-fluid coupling 
given by $-m_b \gamma (\vec{v}_i-\vec{u}_i)$ with 
$\gamma$ the friction coefficient and 
$\vec{v}_{i}$,
$\vec{u}_i$ the bead and fluid velocities at the position 
$\vec{r}_i$ of bead $i$ with a mass $m_b$;
$\vec{F}_{r,i}$ is a random force on bead $i$ with zero mean; 
$\vec{F}_{\kappa,i}$ is the reaction 
force resulting from $N_0-1$ holonomic constraints for molecules modelled 
with rigid covalent bonds, with $N_0$ the number of beads in the 
polymer; and 
$\vec{F}_{drive,i}$ is the driving force representing the effect of the 
external field $\vec{E}$, equal to $q\vec{E} g(\vec{r}_i)$ 
with $q$ an effective charge, which acts only on beads in the pore region.
The region over which the external field acts is described by the
function $g(\vec{r}_i)$, which is
$1$ for $\vec{r}_i$ within this region and $0$ otherwise;
the extent of this region is chosen to be a cube of side 3$\Delta x$
(shown by the green shaded region in Fig. \ref{LJinteractions}),
where $\Delta x$ is the lattice spacing.

For the bead-bead interaction, other than that between 
consecutive beads, and for the 
bead-wall interaction we choose separately a truncated Lennard-Jones potential
(repulsive part only~\cite{WCA}):
\begin{equation}
V(r) =\left\{ \begin{array}{ll}
4 \varepsilon \left [ \left(\frac{\sigma}{r}\right)^{12} - 
\left(\frac{\sigma}{r}\right)^6 \right ]  & \textrm{if $r\leq r_{cut}$}\\
0 & \textrm{if $r>r_{cut}$} \\
\end{array}\right.
\label{eq:LJ}
\end{equation}
In this expression, $r$ is the bead-bead or bead-wall distance.
In both cases the potentials are truncated at the cut-off 
distance $r_{cut}=2^{1/6} \sigma$. 
The chosen parameters $\sigma$ and  $\varepsilon$ are 
$1.8 ~\Delta x$ and $10^{-4}~\Delta m \Delta x/\Delta t^2$ 
for the bead-bead interactions and  $1.5 ~\Delta x$ and $ 10^{-4}~\Delta m \Delta x/\Delta t^2$ 
for the bead-wall interactions. The parameters are given in LB units as explained in Section III.
The distance  between consecutive beads 
along the chain representing the macromolecule is set to $b=1.2 \; \Delta x$ through a constraint imposed by the 
SHAKE algorithm~\cite{SHAKE}.
The geometry of the pore region is shown schematically in Fig. \ref{LJinteractions}.
The fluid is represented through lattice Boltzmann particles
that reside on a three-dimensional cubic lattice with spacing $\Delta x$. 
The probability 
distribution $f_p(\vec{x},t)$ denotes the number of particles at the lattice 
position given by $\vec{x}$ at time $t$, and 
evolves in space and time toward 
the equilibrium distribution $f_p^{eq}$ and with relaxation 
frequency $\omega$, as: 
\begin{eqnarray} 
\label{LBE}
f_p(\vec{x}+ \vec{c}_p \Delta t,t+\Delta t) &=& f_p(\vec{x},t) - 
\omega \Delta t (f_p-f_p^{eq})(\vec{x},t) \nonumber \\
&&+  F_p \Delta t + G_p \Delta t
\label{lbe2}
\end{eqnarray} 
The LB particles can only move on the
lattice with fixed first- and second-neighbor 
speeds $\vec{c}_p$ (19 in all, for a 3D cubic lattice), while
$F_p$ represents thermal fluctuations and 
$G_p$ describes the polymer-fluid back reaction            
\begin{equation}
G_p (\vec{x},t) = w_p  \beta
\sum_{i \in D(x)} [ \vec{F}_{drag,i} + \vec{F}_{r,i} ] \cdot \vec{c}_p
\label{sterm}
\end{equation}
with $w_p$ a set of weights normalized to unity, 
and $\beta$ the inverse fluid temperature. Finally, $D(x)$ denotes 
the lattice cell to which the $i^{th}$ bead belongs.
Details on the numerical implementation of the scheme 
have been reported in Ref. \cite{ourLBM}.

Compared to other numerical methods, the present lattice Boltzmann-Molecular 
Dynamics (LB-MD) scheme has certain computational advantages, namely, 
it permits to take into account
self-consistent hydrodynamic correlations at computational cost
scaling linearly with the polymer length. This allowed us to simulate
long chains in 3D over large statistical ensembles at an
affordable computational cost. 
We simulate polymers of various sizes and as large sample realizations
as our computational resources permit; specifically we considered
sizes of  $N_0$=20 [1000], $N_0$=50 [1000], $N_0$=100 [500], $N_0$=200 [300],
$N_0$=300 [300], $N_0$=400 [200],  $N_0$=500 [150], where the numbers
in brackets are the sample realizations.

\section{Choice of the simulation parameters}
 
We discuss next the details of the choice of parameters in the model so that our simulations
are relevant to DNA translocation through nano-pores, as observed in typical 
experimental setups~\cite{NANO}.
The simulations are performed in a three-dimensional box
which contains the polymer and the fluid solvent, 
and has a size $N_x \times N_x/2 \times N_x/2$ 
in units of the lattice spacing $\Delta x$; we used 
$N_x=80$ for polymers of size $N_0\leq400$ beads.
The separating wall is located in the mid-section of the $x$ direction,
at $x/\Delta x= N_x/2$, with a square hole of side $3\, \Delta x$ 
at the center through which the polymer translocates from 
one chamber to the other. At $t=0$ the polymer resides entirely 
in the right chamber at $x/\Delta x>  N_x/2$, with its one end 
at the pore region 
along the streamline ($x$) and cross-flow ($y,z$) directions. 
The LB time step is $\Delta t$ and the MD time step $\Delta t/5$.

In order to have a plausible representation of DNA, we choose the separation between
consecutive beads to be $b = 50$ nm, the persistence length of double-stranded DNA.
Since we defined $b = 1.2 \; \Delta x$, this gives for the lattice spacing $\Delta x = 42$ nm.
Since the repulsive interaction between the beads and the wall 
involved the parameter $\sigma_w=1.5~\Delta x$, and the pore is a square of size $3 \Delta x$,  
this produces an effective hole of size $\sim 5$ nm through which the polymer 
translocates.
Having set the value of $\Delta x$, 
we choose the time step so that the kinematic viscosity is expressed as:
$$\nu_w = \nu_{LB} \frac{(\Delta x)^2}{\Delta t},$$ 
with $\nu_w$ the viscosity of water
($ 10^{-6}$ m$^2$/s) and $\nu_{LB}$ the numerical value of the viscosity in LB units;
taking $\nu_{LB}=0.1$ produces a time-step of  $\Delta t = 160$ ps.

The above choice for the lattice spacing and time-step
fixes the lattice speed $c \equiv (\Delta x/\Delta t) \sim 250$ m/s, which is reasonably
close to the solvent thermal speed $\sqrt {kT/m_w} \sim 500$ m/s.
It is instructive to compare the lattice speed, $c$, with the typical propagation speeds
of the main phenomena of interest in the translocation process.
These are: \\
i) the translocation speed $ b (dN/dt)  \sim \; 10^{-3}$ m/s; \\ 
ii) the electric drift speed $ (qE)/(m_{bp} \gamma_{bp}) \sim 0.3$ m/s; \\ 
iii) the base-pair thermal speed $ \sqrt {kT/m_{bp}} \sim 70$ m/s,\\
where we have used the following reference values for the base-pair translocation rate
$dN/dt \sim 10^7$ bp/s,
the strength 
of the force due to the external field 
$qE \sim 10$ pN,  
the mass of a base-pair $m_{bp}$ in terms of the 
mass of the water molecule $m_w$, 
$m_{bp} \sim 30 \;, m_w \sim 600$ amu, 
the drag coefficient for a base-pair
 $\gamma_{bp} = 3 \times 10^{13}$ s$^{-1}$~\cite{LU}  and 
$T=300 \;^0$K for the temperature.
These order-of-magnitude estimates 
indicate that the present choice of space and time units is such that 
the corresponding speed, $u$,
fulfills the numerical stability CFL (Courant-Friedrichs-Lewy) condition, $u<c$.

In LB simulations mass units are fixed by the mass density of the fluid species $\rho_w$.
Setting the LB mass density $\rho_{LB}=1.0$ corresponds to  
having a number of water molecules $\rho_w\Delta x^3/\rho_{LB} m_w = 2\times 10^6$,
where $\rho_w$ is the density of water, and each lattice site contains a solvent
mass $\Delta m = \rho_w \Delta x^3 /\rho_{LB}$.
Since we are using a continuum-kinetic representation of fluid flow at the nanoscale, 
a necessary condition for this representation to hold against
statistical noise is that the above ratio be much greater than unity.
Given the fact that we have $19$ discrete distributions, $f_i$ per cell, each of them 
would represent about $10^5$ water molecules, a safe value towards satisfying this
condition.

The prime goal of the mapping procedure is to
secure the correct values of the major dimensionless parameters governing the physics
of the translocation process. 
In particular, this regards the ratio of external drive to thermal forces which we will call $\phi$, 
defined as
\begin{equation}
 \phi \equiv \frac{qEb}{kT}
 \label{eq:phi-definition}
\end{equation}
The value of this quantity 
in actual experiments is $\phi \sim 1-10$ \cite{NANO}.  In our simulations, we took $kT=10^{-4}$ and
$qE = 0.01$ acting on one bead, mapping $\simeq 100$ base-pairs. 
These quantities are again given in LB units. At a base-pair level, 
this means $qE=10^{-4}$, corresponding to $\phi = 1$, in satisfactory
order-of-magnitude agreement with experiments. With this value of the driving force and the
polymers studied in the present work,
a typical translocation event takes place in a time interval on the order of $10^3-10^4$ LB timesteps.

In the simulations we define the effective mass of the beads to be $m_b=1$. 
A straightforward calculation shows that the ratio
of the effective bead mass to the mass resulting from the DNA coarse graining, that is, one 
bead representing $\sim 100$ base pairs, is $\sim 700$.
On the other hand, the parameter relevant to momentum exchange is the bead friction $\gamma$.
We chose $\gamma =0.1$ in LB units, so that the ratio between friction in physical units
and the experimentally determined one ($\gamma_{bp}=30$ ps$^{-1}$ ~\cite{LU}) 
is $10^{-4}$. Our choice was dictated by the criterion of numerical stability 
$\gamma < 1/\Delta t = 6\times 10^{9}$ s$^{-1}$.
These factors taken together show that the bare particle mobility, $\mu=1/(m_b\gamma)$, 
is a factor of $15$ larger than the experimental one.
This is equivalent to an underdamped motion of the macromolecule, which 
results in smooth particle trajectories and allows for algorithmic 
stability without affecting the long-time behavior of the polymer.
A possible alternative would be to solve the polymer dynamics in overdamped (Brownian) form
which also circumvents the small timestep issue imposed by the frictional damping 
\cite{ottingerbook,ottingerthanks}. 
Our choice to use inertial dynamics was based on previous experience with similar
systems, without hydrodynamic interactions, which showed that the inertial dynamics
approach has a slight advantage in stability with larger time-step size.

\section{Translocation simulations}

\begin{figure}
\begin{center}
\includegraphics[width=0.5\textwidth]{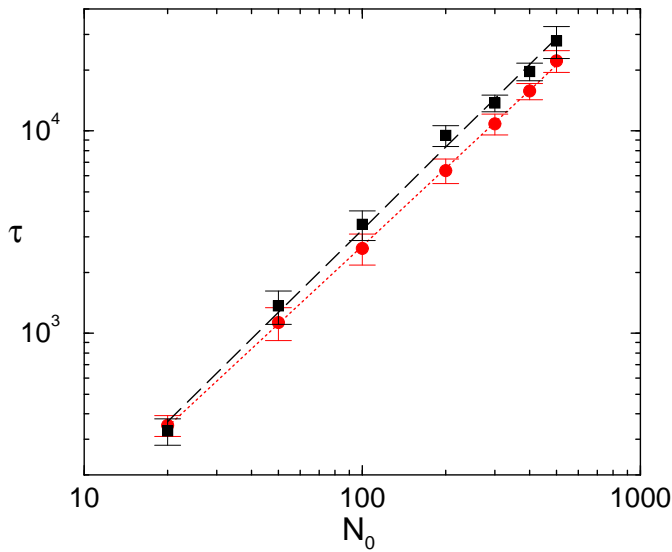}
\caption{(Color online) Scaling of the translocation time $\tau$ with the
number of beads $N_0$, in the presence (circles) or absence 
(squares) of coupling between the molecule and the solvent. 
The exponents are $1.28\pm0.01$ and 
$1.36\pm0.03$, respectively.}
\label{FIG1}
\end{center}
\end{figure}

We report next the results of extensive simulations of translocation events.
In each case, the translocation time exhibits a statistical (not 
exactly gaussian) distribution around the most probable translocation 
time, in close analogy to experimental observations 
\cite{EXPRM1,EXPRM2,NANO}. 
 According to these studies, a 48.5 kbp double-stranded DNA takes
about $2$ msec to translocate, while a typical Zimm time is about $700$ msec. 
Fig.~\ref{FIG1} shows 
the dependence of the most probable translocation time $\tau$ on the 
polymer length, which 
obeys a superlinear scaling relation: $\tau  \propto N^{\alpha}_0$.
The observed exponent, with the molecule-fluid coupling in operation, 
$\alpha \simeq 1.28\pm0.01$ is in very good agreement with 
recent experiments of DNA translocation through 
a nanopore driven by an external electric field, 
where $\alpha \simeq 1.27\pm0.03$ \cite{NANO}. 

In the absence of coupling between the molecule and the solvent,
the translocation process is 
slowed down, as indicated by a higher
exponent, $\alpha \simeq 1.36\pm0.03$.
We propose that in the presence of polymer-solvent coupling
some form of bead-bead screening takes place, 
the biopolymer never translocates in the form of a linear chain, 
as this configuration is entropically suppressed. 
The physics of the larger exponent (slower motion) in the absence of 
hydrodynamics is related to enhanced correlation effects and 
a wider range of polymer fluctuations.

\begin{figure}
\begin{center}
\includegraphics[width=0.5\textwidth]{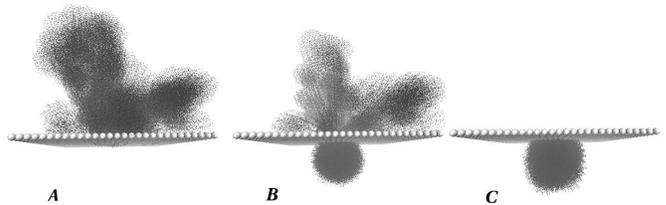}
\caption{3-D view of an ensemble of $100$ polymers with $N_0=300$ 
at different stages of the 
process: $A$, $B$ and $C$ correspond to the 
initial, mid-point, and final translocation times.}
\label{MEDUSA} 
\end{center}
\end{figure}

\begin{figure}
\begin{center}
\includegraphics[width=0.45\textwidth]{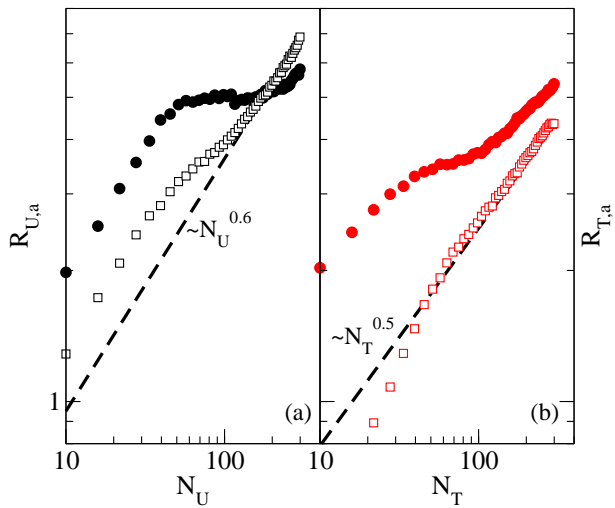}
\caption{(Color online) 
Longitudinal (filled circles) and transverse (open squares) components of the 
gyration tensor for (a) the $U$ and (b) the $T$ segment
with the number of untranslocated (N$_U$) and translocated (N$_T$) beads, respectively.
The dashed lines show the scaling.}
\label{ANISO} 
\end{center}
\end{figure}

In Fig. \ref{MEDUSA} we show an ensemble of $100$ polymer configurations
at three different instants, referring to the initial, mid-point and final
stages of translocation.
From this figure it is clear that, initially, the shape of the untranslocated (U) segment
is squeezed against the wall and takes the form of an oblate ellipsoid.
The translocated (T) segment appears to be more compact and prolate. 
The tendency of the U-segment to be attracted to the wall is in line with
the well-known ``mushroom'' shape arising from a polymer anchored at one
end to a repulsive wall \cite{degennes,binder}.
To investigate the consequences of this anisotropy, we have
inspected the scaling of the gyration tensor with the number of 
monomers for the U and T segments separately, 
with time, according to:
\begin{equation}
R_{I,\mu}(t) \propto  [N_{I}(t)]^{\nu_{I,\mu}}, (I=U,T; \; \; \mu=\perp,\parallel)
\label{r_scaling}
\end{equation}
where $\mu = \parallel, \perp$ denotes the longitudinal 
and transverse components of the gyration tensor, with respect to 
the direction of translocation.
This scaling is shown in Fig. \ref{ANISO},
from which we observe that for $N_U > 100$ the transverse component
$R_{U,\perp}$ follows a dynamic scaling law with
$\nu_{U,\perp} \simeq 0.6$, close to the Flory-exponent of a 
3-D self-avoiding random walk, while
the longitudinal one exhibits a much weaker dependence 
on $N_U$ (smaller slope).  The T-segment follows a similar trend, but with
a transverse component scaling with $\nu_{T,\perp} \simeq 0.5$.

The fluid/biopolymer system can be approached as an extended dynamical 
system consisting of two components: this system receives energy from 
the exterior through a localized electric field acting on the polymer 
and dissipates it via interaction of both the fluid and the polymer 
with the wall. Each bead is subject to the following 
forces: (a) the localized drive $\vec{F}_{drive,i}$, (b) the dissipative 
drag $\vec{F}_{drag,i}$, (c) the pore drag $\vec{F}_{pore,i}$, and  
(d) the entropic forces. At equilibrium, the latter can be expressed as 
\begin{equation}
F_{entr} \sim  \frac{k_BT}{b} \frac{1-2r}{r(1-r)}
\label{eq:entropic}
\end{equation} 
where $r(t) \equiv N_T(t)/N_0$ is the translocation coordinate \cite{Nelson} and
$b$  is the separation between two beads.
Here, we have used an explicit dependence on $r$ simply to show that entropic
forces are negligible most of the time, except at both ends (initiation and completion) 
of the translocation process. 
These forces are naturally measured in terms of the thermal force,
$F_{th}=k_BT/b$. In the fast translocation regime considered 
in this work, $F_{drive}/F_{th} >1$, yielding $bF_{drive}/k_BT \sim 10^2$
for the parameters used.
Pore forces are negligible in our simulations,
due to the small pore size. 
For the entropic forces,
$F_{entr}/F_{th} \sim (1-2r)/(r(1-r))$, 
which shows that they can be neglected except at 
the early ($r(t) \to 0$) and final ($r(t) \to 1$) stages of the 
process. With entropic forces and pore dissipation 
negligible, the forces guiding the translocation
are the hydrodynamic drag and the drive from the external electric field.

Hydrodynamics is expected to provide
a cooperative background, helping to minimize frictional effects. 
For quantitative insight into this, we monitor the {\it synergy} factor, 
defined as the work per unit time made by the fluid on 
both parts ($I$=$U$ or $T$) of the polymer:
\begin{equation}
S_{H}^{(I)}(t)=
\frac{dW_{H}^{(I)}}{dt}=\gamma \Big\langle\sum_i^{N_I}  
\vec{v}^{(I)}_i(t) \cdot \vec{u}_i(t)\Big\rangle
\label{eq:Shydro}
\end{equation}
where brackets denote averages over different realizations
of the polymer for a given length.
Positive (negative) values of $S_{H}^{(I)}(t)$ 
indicate a cooperative (competitive) solvent, respectively.
The variation of $S_{H}^{(I)}(t)$ with time is linear
(Fig.~\ref{FIG4}(a)), while the total rate
$S_H(t)=S_{H}^{(T)}(t)+S_{H}^{(U)}(t)$ on the whole chain, 
is constant with time indicating that the work per unit time
associated with the change of the radii of the two blobs 
is constant.
The probability distribution of $S_{H}(t)$ during translocation
(Fig.~\ref{FIG4}(b)), indicates that hydrodynamics 
turns the solvent into a cooperative environment: 
the distribution lies entirely in the positive range. 
This cooperative effect is the underlying reason for the faster 
translocation process in the presence of hydrodynamic interactions.
Similarly, the work done per timestep by the electric field on the 
polymer can be defined as:
\begin{equation}
S_E (t)=\frac{dW_E}{dt} = 
\Big\langle\sum_i \vec{F}_{drive,i} \cdot \vec{v}_{i}(t) 
\Big\rangle
\label{eq:Selec}
\end{equation}
The average of $S_E(t)$ 
is positive (Fig.~\ref{FIG4}(a)).
However, the negative tail of the corresponding probability
distribution in Fig.~\ref{FIG4}(b) indicates that there is a non-negligible
probability to find beads moving against the electric field. 
On average, $S_E(t)$ is also constant with time (except very near the 
completion of the translocation) denoting that
the beads traverse the pore with basically the same speed at all times.
The average hydrodynamic work per time is larger
than the input of the external electric field, since the latter only acts
on a very small fraction of the beads, about $4$ resident beads within 
the pore region.

\section{Phenomenological model}

We next present a phenomenological model for the scaling exponent
of the translocation time in the presence of the solvent.
We first consider Eq.(\ref{eq:force}) and multiply both sides of the equation
with the bead velocity $\vec{v}_{i}$. 
Summing up over all beads, using the fact that the velocities 
are uncorrelated with the random forces $\vec{F}_{r,i}$
and that $\vec{F}_{\kappa,i}$ produces no work, we obtain:
\begin{equation}
\frac{dK}{dt}= 
\left(\frac{dW}{dt}\right)^{(1)} +
\left(\frac{dW}{dt}\right)^{(2)} +
S_{E}(t)
\label{eq:dwdtSum}
\end{equation}
where $K$ is the kinetic energy of the polymer and 
the right-hand side of  this equation
contains the change in energy of the U and T sections 
of the polymer due to the mechanical work $W^{(1)}$, the viscous drag,
$W^{(2)} = S_H - 2 \gamma K$, and the rate of work done by the electric field, $S_E$.
By writing Eq. (\ref{eq:dwdtSum}) we assume that
the translocation time $\tau(N_0)$  for the entire polymer is determined by two
separate contributions. 
The first stems from the change in free energy of the polymer due to the
removal of beads from the U side and their addition to the T side. 
The second term arises from shifting the center of mass of each blob from the initial
position towards (for the U part), or away from (for the T part), the pore entrance. 

\begin{figure}
\begin{center}
\includegraphics[width=0.45\textwidth]{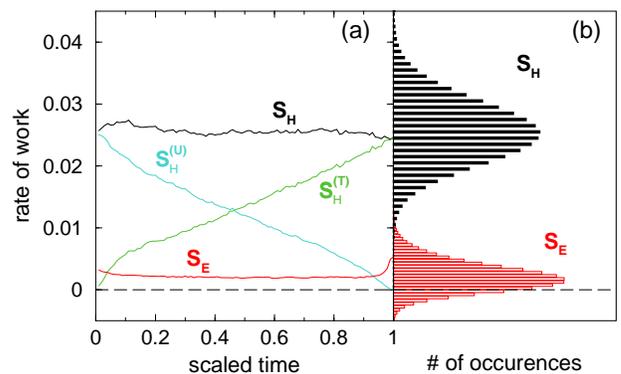}
\caption{(Color online) (a) Synergy factors with time for the hydrodynamic 
($S_{H}^{(T)}$, $S_{H}^{(U)}$,
total $S_{H}$) and electric 
field ($S_{E}$). (b) Probability distributions of $S_{H},~S_{E}$
during translocation events. 
Curves are averages over 100 events for $N_0=300$.}
\label{FIG4}
\end{center}
\end{figure}

The simulations reveal that
$S_H$ and $S_E$ are each independently constant in time to a very good approximation. 
This holds for both the average values over all samples
(see Fig.~\ref{FIG4}(a)) and for any polymer realization.
The average $K$ (not shown) is also
approximately constant, as the temperature is constant, leading to $dK/dt=0$.

Regarding the mechanical work $W^{(1)}$, simulation data show that,
except for a short-lived transient at the beginning and final part of the process, the
rate of removal/addition of beads is linear in time. Similarly, 
for the viscous drag the shift of the center of mass
takes place at constant velocity, as illustrated in Fig.~\ref{FIG4} by the two individual 
components $S^{(U)}_H(t)$  and $S^{(T)}_H(t)$, both of which show linear behavior. 
Since the two contributions can be viewed as independent components of
the work, we separately analyze their effects on the scaling dependence.

In order to estimate the work required to shrink/grow the two blobs, we now introduce a macroscopic 
picture according to which, for a spherical blob of radius $R_I$, surface $A_I$ and volume $V_I$
the work is $dW_I^{(1)}=P_IdV_I+\sigma_{\gamma}dA_I$ with 
$\sigma_{\gamma}$ the average surface tension
and $P_I=2\sigma_{\gamma}/R_I$ the pressure acting on the blob 
(Laplace equation).
Given the anisotropy of the $I = U$ and $T$ segments, 
the above relation generalizes to
$dW^{(1)} = \sum_I \sum_{\mu=\perp,\parallel} \lambda_{I,\mu} R_{I,\mu}dR_{I,\mu}$
with $\lambda_{I,\mu}$ collecting all constants. 
More explicitly, 
$R_{U,\perp} dR_{U,\perp} \sim N_U^{2\nu_U-1} dN_U$ and 
$R_{T,\perp} dR_{T,\perp} \sim N_T^{2\nu_T-1} dN_T$, 
with $\nu_U \simeq 0.6$ and $\nu_T \simeq 0.5$, where 
the longitudinal components of both U and T segments have been neglected
in view of their much weaker dependence on the number of beads.
The rate of work on the entire polymer, consisting of the 
$U$ and $T$ blobs with radii given by Eq.(\ref{r_scaling}),
takes then the form
\begin{eqnarray}
\frac{dW^{(1)}}{dt} &=&   
\left[\lambda_{T,\perp}^2 N_0^{2 \nu_T} r^{2\nu_T-1} \frac{dr}{dt}
     -\lambda_{U,\perp}^2 N_0^{2 \nu_U} (1-r)^{2\nu_U -1} \frac{dr}{dt} \right] 
\label{eq_rate}
\end{eqnarray}
with $r=N_T(t)/N_0$. Since $dW^{(1)}/dt$ is constant, integration of the above 
expression, with $r\in[0,1]$ and $t\in[0,\tau]$, 
leads to the scaling of the total translocation time:
$\tau \sim  N_0^{2\nu_U}$. 
It is worth mentioning that, at variance with a previous argument \cite{NANO}, our macroscopic picture
does not require that each blob be in a state of mechanical equilibrium.
In fact, it is clear from Fig. \ref{MEDUSA} that at the end of translocation the blob
is definitely not in equilibrium.

We next consider the viscous drag due to the net motion of the blob relative to the fluid,
which can be computed by the global friction
experienced by the whole set of $N_0$ monomers
$F_{drag} = \sum_{i=1}^{N_0} \gamma (u_i-v_i)$, where again $v_i$ is the bead
velocity, $u_i$ is the fluid velocity at the monomer location and $\gamma$
is the friction coefficient.
This is best recast in the form of an effective friction coefficient
$F_{drag} = -\gamma_{eff} {\cal V}$, where ${\cal V}$ is the center of mass velocity of the blob.  
As is well known \cite{doi}, in the absence of hydrodynamic
correlations all monomers behave independently, so that $\gamma_{eff}$ scales like $N_0$. 
On the other hand, when hydrodynamic correlations are included, the inner monomers are
screened out from the outer ones, so that the effective friction is reduced
and scales less than linearly with length. More precisely, $\gamma_{eff} \sim R \sim N_{0}^{\nu}$.
Given that $ {\cal V}=dR/dt$, the hydrodynamic drag 
scales like $N_{0}^{1+\nu}$ and $N_{0}^{2\nu}$, without and with 
hydrodynamics, respectively.  
Remarkably, the exponent with hydrodynamics, $2 \nu$, is exactly the same as the one associated with
the thermodynamic work $W^{(1)}$, so that $\alpha = 2 \nu$ in either case.
In the case without hydrodynamics, however, the thermodynamic work and the work due to viscous drag exhibit two
distinct exponents, $2 \nu\simeq 1.2$ and $1+\nu\simeq 1.6$, which explains why any attempt to represent
the scaling through a single exponent $\zeta$ is bound to work only on a narrow range
of values of $N_0$. Generally, the scaling $\tau(N_0)\sim N_0^\zeta$ will be a weighted average of the two,
i.e. $2\nu < \zeta < 1 + \nu$.

In support of the previous interpretation, we have measured the typical values of the rate of change
of the blob radius $\dot R_I$, the mean center-mass velocity ${\cal V}$ and the average flow speed ${\cal U}$
and found that $\dot R_I \simeq {\cal V}_I \simeq 5 {\cal U}_I$. This corroborates the idea put forward 
in this paper according to which both mechanisms, the blob shrinking as well as the blob shifting processes,
must be taken into account to provide a complete picture.

\section{Discussion and Conclusions}

Summarizing, we have investigated the process of polymer 
translocation through a narrow pore using a multiscale approach which
explicitly accounts for the hydrodynamic interactions of the 
molecule with the surrounding solvent. 
The translocation time was found to obey a power-law dependence on the
polymer length, with an exponent $\alpha = 1.28 \pm 0.01$, 
in a satisfactory agreement with experimental measurements
and other computer simulations.
Moreover, our simulations reveal that the coupling of the
molecular motion to hydrodynamic correlations results in 
a significant acceleration of the translocation process.
The scaling behavior observed in the numerical simulations has
been interpreted by means of a new
phenomenological model, accounting for the anisotropy 
of both translocated and untranslocated segments.
This ingredient appears to be crucial to the
correct interpretation of the basic mechanisms behind the
physics of the polymer translocation, which involves two
separate processes, the shrinking and the shifting of the blob.

Deviations from the mean-field picture
occur mainly near completion of the process, where the
radius of the untranslocated segment undergoes an accelerated depletion:
for $r(t)>1/2$, the majority of the beads have already translocated and
entropic forces cooperate with the electric field to complete
the translocation.
Besides violating the static scaling at the end of the 
translocation process, entropic forces may lead to
more dramatic effects, which escape any mean-field description  
based on the translocation coordinate $r(t)$ alone.  
Such beyond-mean-field-theory effects produce rare retraction events: 
the polymer occasionaly anti-translocates after having partially passed through the pore. 
Our simulations reveal that retraction events are typically 
associated with the $T$
part entering a low-entropy (hairpin-like) 
configuration, which is then subject to a strong entropic pull-back.
These non-perturbative events depend on the polymer length, 
the initial configuration and the values of 
other parameters (friction 
constant, temperature and strength of the pulling force).
They occur at a rate up to 2\% and 
do not significantly affect the statistics 
of the scaling exponent. 

\acknowledgments
MF acknowledges support by Harvard's Nanoscale Science and 
Engineering Center, funded by NSF (Award No. PHY-0117795).
SM and SS wish to thank the Physics Department at Harvard
University for kind hospitality.
We wish to thank H.C. \"Ottinger for valuable discussions.

\end{document}